\documentclass[11pt]{article}
\usepackage{graphicx,amsmath,amssymb}
\usepackage{graphics}
\usepackage{color}
\usepackage[colorlinks,
linkcolor=black,
anchorcolor=black,
citecolor=black,
urlcolor=black
]{hyperref}

\usepackage{breakurl}

\usepackage{url}

\makeatletter
\newcommand{\rmnum}[1]{\romannumeral #1}
\newcommand{\Rmnum}[1]{\expandafter\@slowromancap\romannumeral #1@}
\makeatother

\begin{document}

\newtheorem{definition}{Definition}[section]
\newtheorem{theorem}{Theorem}[section]
\newtheorem{lemma}[theorem]{Lemma}
\newtheorem{corollary}[theorem]{Corollary}
\newtheorem{proposition}[theorem]{Proposition}

\newcommand{\blackslug}{\penalty 1000\hbox{
    \vrule height 8pt width .4pt\hskip -.4pt
    \vbox{\hrule width 8pt height .4pt\vskip -.4pt
          \vskip 8pt
      \vskip -.4pt\hrule width 8pt height .4pt}
    \hskip -3.9pt
    \vrule height 8pt width .4pt}}
\newcommand{\proofend}{\quad\blackslug}

\newenvironment{proof}{\medskip\noindent {\sc Proof.}$\;\;\;$\rm}{\qed}

\newcommand{\qed}{\hspace*{\fill}\blackslug}

\def\boxit#1{\vbox{\hrule\hbox{\vrule\kern4pt
  \vbox{\kern1pt#1\kern1pt}
\kern2pt\vrule}\hrule}}

\addtolength{\baselineskip}{+0.4mm}

\title{\bf Frameworks for Solving Turing Kernel Lower Bound Problem and Finding Natural Candidate Problems in NP-intermediate}

\author{{Weidong Luo}\thanks{Central South University (My Alma Mater), Changsha, P.R. China. E-mail addresses: luoweidong@csu.edu.cn or weidong.luo@yahoo.com}\\}
\date{}
\maketitle

\vspace{-7mm}

\begin{abstract}
Kernelization is a significant topic in parameterized complexity. Turing kernelization is a general form of kernelization. In the aspect of kernelization, an impressive hardness theory has been established [Bodlaender etc. (\href{http://link.springer.com/chapter/10.1007\%2F978-3-540-70575-8_46}{ICALP 2008}, \href{http://www.sciencedirect.com/science/article/pii/S0022000009000282}{JCSS2009}), Fortnow and Santhanam (\href{http://dl.acm.org/citation.cfm?doid=1374376.1374398}{STOC 2008}, \href{http://www.sciencedirect.com/science/article/pii/S0022000010000917}{JCSS 2011}), Dell and van Melkebeek (\href{http://dl.acm.org/citation.cfm?doid=1806689.1806725}{STOC 2010}, \href{http://dl.acm.org/citation.cfm?doid=2629620}{J. ACM 2014}), Drucker (\href{http://ieeexplore.ieee.org/document/6375340/}{FOCS 2012}, \href{http://epubs.siam.org/doi/10.1137/130927115}{SIAM J. Comput. 2015})] based on the assumption that polynomial hierarchy will not collapse, and this hardness theory can obtain lower bounds of kernel size. Unfortunately, this framework is not fit for Turing kernelization. Moreover, so far, there is no tool which can be applied to obtain Turing kernel lower bound for any FPT problem modulo any reasonable complexity hypothesis. Thus, constructing a framework for lower bounds of Turing kernels of FPT problems has received much attention from the parameterized complexity community, and this has been proposed as an open problem in different occasions [Fernau etc. (\href{http://drops.dagstuhl.de/opus/volltexte/2009/1843/}{STACS 2009}), Misra etc. (\href{http://www.sciencedirect.com/science/article/pii/S157252861000068X}{Discrete Optimization 2011}), Kratsch (\href{http://citeseerx.ist.psu.edu/viewdoc/summary?doi=10.1.1.673.3999}{Bulletin of the EATCS 2014}), Cygan etc. (\href{http://fptschool.mimuw.edu.pl/opl.pdf}{Dagstuhl Seminars on kernels 2014)}].

Ladner [\href{http://dl.acm.org/citation.cfm?doid=321864.321877}{J. ACM 1975}] proved that if P $\not=$ NP, then there exist infinitely many NP-intermediate problems. However, the NP-intermediate problem constructed by Ladner is artificial. Thus, finding natural NP-intermediate problems under the assumption of P $\not=$ NP interests theorists, and it has been a longstanding open problem in computational complexity community.

This paper builds a new bridge between parameterized complexity and classic computational complexity. By using this new connection, some frameworks can be constructed. Based on the assumption that the polynomial hierarchy and the exponential hierarchy will not collapse, these frameworks have three main applications. Firstly, these frameworks can be used to obtain Turing kernel lower bounds of some important FPT problems, thus solving the first open problem. Secondly, these frameworks can also be used to obtain better kernel lower bounds for these problems. Thirdly, these frameworks can be used to figure out a large number of natural problems in NP-intermediate, thus making some contributions to the second open problem.
\end{abstract}

\newpage
\section{Introduction}
Parameterized complexity extends classical complexity theory by considering not only the input length but also one or more additional parameters like solution size or structural measures like treewidth. A parameterized problem is a language $L \subseteq \Sigma^* \times \mathbb{N}$, where $\Sigma$ is an alphabet with two or more symbols, and the second component is called the parameter of the problem. Kernelization is a well-known and important topic in parameterized complexity \cite{cygan book, downey book, guo book}. It is a theoretical formalization of efficient preprocessing to deal with hard languages. The polynomial time preprocessing, which is a kernelization, is used to shrink the size of instance and output an equivalent instance, and the output of the preprocessing is called kernel. Another significant topic in parameterized complexity is fixed-parameter tractable. A parameterized language $L$ is fixed-parameter tractable if there exists an algorithm to decide whether $(x,k) \in L$ in time $f(k)|x|^{O(1)}$, where $f$ is an arbitrary function depending only on $k$, and the corresponding complexity class is called FPT. There is an important connection between kernelizaiton and FPT, that is, a problem has kernel if and only if the problem is in FPT.

Unless otherwise stated, we will use following definition in this paper. Functions $f: \mathbb{N} \rightarrow \mathbb{N}$, $g: \mathbb{N} \rightarrow \mathbb{N}$, $t: \mathbb{N} \rightarrow \mathbb{N}$ are nondecreasing. $F$ and $T$ are classes of functions.
\begin{definition}
 (Kernelization) A kernelization for a parameterized language $L \subseteq \Sigma^* \times \mathbb{N}$ is a polynomial-time Turing machine $M$ that given any instance $(x,k)$ returns an instance $(x',k')$ such that $(x,k)\in L$ if and only if $(x',k')\in L$ and with $|x'|,k' \leqslant f(k)$. We also call the kernelization as $f$-sized kernelization.
\end{definition}
We say $L$ has an $F$-sized kernelization if $L$ has an $f$-sized kernelization for some $f \in F$. For example, if $f(k)$ is polynomially (linearly) bounded in $k$, then $M$ is a polynomial (linearly) kernelization and we say the language $L$ has a polynomial (linearly) kernel.

Kernelization requires that the input and output instances belong to the same language. Without this requirement, the concept of kernelization transforms into the concept of compression.
\begin{definition}
(Compression) A compression for a parameterized language $L\subseteq \Sigma^* \times \mathbb{N}$ is a polynomial-time Turing machine $M$ that given any instance $(x,k)$ returns an instance $x'$ such that $(x,k) \in L$ if and only if $x' \in Q$ ($Q$ is a language) and with $|x'|\leq 2f(k)$. We also call the compression as $f$-sized compression.
\end{definition}
The way we define $F$-sized kernelization can be also used to define $F$-sized compression. Compression is a general type of kernelization. In some languages, we can probably shrink the instances into smaller size with compression rather than kernelization. For example, Wahlstr\"{o}m \cite{tutte matrix} proved that $K$-cycle problem has a $|K|^3$-sized compression, but a polynomial kernel has not been found so far.

As more and more techniques for kernelization were found. People realized that it was hard for some problems, such as $k$-path problem, which asks whether a graph has a simple path of length $k$, to obtain small kernels, such as polynomial kernel, by using existing techniques. Thus, to figure out why these problems are so hard to get small kernels, researchers proposed an impressive hardness theory to prove lower bounds for kernel size. In their paper \cite{Bodlaender1}, Bodlaender etc. firstly raised two classes of problems. Then, they proved that the problems in one class (including $k$-path) have no polynomial kernels, unless or(SAT) can compress, and that the problems in another class have no polynomial kernels, unless and(SAT) can compress. Fortnow and Santhanam \cite{fortnow1} proved that or(SAT) can not be compressed, unless co-NP $\subseteq$ NP/poly and the polynomial hierarchy collapse. However, and(SAT) had remained an open problem for five years until it was settled by Drucker \cite{Drucker}, based on the assumption that there is no nonuniform, statistical zero-knowledge proofs for NP languages, which is a weaker one compared with co-NP $\not \subseteq$ NP/poly. Another insightful work of kernel lower bounds is attributed to Dell and van Melkebeek \cite{van Melkebeek}. They gave a framework for proving kernel lower bounds for problems that do admit some polynomial kernelization. For example, they showed that d-SAT has no $O(n^{d-\epsilon})$ kernel unless co-NP $\subseteq$ NP/poly and polynomial hierarchy collapse. Dell and Marx \cite{dell98}, Hermelin and Wu \cite{hermelin} and Bodlaender etc. \cite{a13} also made contributions to the hardness theory. Since then, a large amount of researches have been conducted to solve various concrete problems by employing these frameworks \cite{ a7, a5, a1, cygan3456, a6, a4, hermelin, a3, a11, a10, a9, a2, a12, a8}. In fact, all of the kernelization lower bound frameworks mentioned above are also fit for compression lower bound. Compression (kernelization) lower bound also means that the smallest size of the output instance can be obtained by compression (kernelization).

However, there are some relaxed notions of kernelization (including Turing kernelization) whose lower bound are not ruled out by any existing techniques, and these relaxed notions had been proposed \cite{Estivill-Castro4534, guo book} before the kernelization lower bound framework was raised.

The relaxed notions allow the kernelization algorithm to output many small independent instances. For example, Fernau etc. \cite{fernau123} found the first problem, namely leaf out-tree($k$), which has no polynomial kernel unless co-NP $\subseteq$ NP/poly. However, they constructed a polynomial time algorithm for the problem whose input is an instance of leaf out-tree($k$), whose output is $n$ independent kernels of $O(k^3)$-sized, and where the input is a yes instance if and only if at least one output instance is yes instance ($n$ is the vertices number of the input instance and $k$ is the parameter). Because the status of the input instance is equivalent to the disjunction of the status of the output kernels, this polynomial time algorithm is called disjunctive kernelization, which is a type of relaxed notion of kernelization, and the output of this polynomial time algorithm is called disjunctive kernel. With the help of the disjunctive kernelization, we can design an algorithm for the original problem by solving each kernel in turn, and this algorithm only needs to deal with at most $n$ independent $O(k^3)$-sized kernels. Therefore, this kind of relaxed kernel is still useful in practice.

According to the definition of disjunctive kernelization, it is easy to understand other relaxed notions of kernelizations, including conjunctive kernelization, truth-table kernelization and Turing kernelization. The most and least powerful of these three relaxed notions are Turing kernelization and conjunctive kernelization, respectively. In this paper, we mainly discuss the most general relaxed notion of kernelization, i.e. the Turing kernelization. The framework which is fit for Turing kernel lower bounds can also be suitable for other types of kernel lower bounds. We recall the definition of Turing kernelization given by \cite{fernau123} below.

\begin{definition}
($f$-oracle for parameterized language) An $f$-oracle for a parameterized language $Q$ is an oracle that takes as input $(x,k)$ with $|x|\leq f(n)$, $k\leq f(n)$ ($n$ is some integer) and decides whether $(x,k) \in Q$ in constant time.
\end{definition}

\begin{definition}
(Turing Kernelization) A parameterized language $L \subseteq \Sigma^* \times \mathbb{N}$ is said to have an $f$-sized Turing kernelization if there is a Turing machine which given an input $(x,k)$ together with an $f(k)$-oracle for $L$ decides whether $(x,k) \in L$ in time polynomial in $|x|+k$.
\end{definition}
We say $L$ has an $F$-sized Turing kernelization if $L$ has an $f$-sized Turing kernelization for some $f \in F$. Like the way define compression, we define Turing compression in the same way by allowing the oracle queries to be any other languages. We will define these variants below.

\begin{definition}
($f$-oracle for classic language) An $f$-oracle for a classic language $Q$ is an oracle that takes as input $x$ with $|x|\leq 2f(n)$ ($n$ is some integer) and decides whether $x \in Q$ in constant time.
\end{definition}

\begin{definition}
(Turing Compression) A parameterized language $L \subseteq \Sigma^* \times \mathbb{N}$ is said to have an $f$-sized Turing compression if there is a Turing machine which given an input $(x,k)$ together with an $f(k)$-oracle for a classic language $Q$ decides whether $(x,k) \in L$ in time polynomial in $|x|+k$.
\end{definition}
We define $F$-sized Turing compression according to the way we define $F$-sized Turing kernelization.

Turing kernelization is more powerful than other relaxed notions of kernelizations. It could even adaptively create reduced instances. So far, people have found some problems that have polynomial Turing kernels, but do not have polynomial kernels \cite{Abhimanyu, a13, fernau123, Jansen, Alexander, Thomass45}. Finding framework for lower bounds of Turing kernels has become an important open problem in parameterized complexity community, and it has been proposed as an open problem in different occasions \cite{open2, fernau123, open1, open3}. In particular, during Dagstuhl Seminars on Kernels in 2014, Cygan etc. \cite{open3} gave a list of open problems and ranked them according to their importance and possible hardness, and this problem is one of the most important and difficult problems in the list. Unfortunately, Unlike lower bounds of kernels, until now, there is no technique that can deal with lower bounds of Turing kernels (even for truth-table kernels and disjunctive kernels) for any FPT problem modulo any reasonable complexity hypothesis. The positive aspect is that the polynomial conjunctive kernels can be refuted by modifying the framework of the lower bounds of many-one kernels \cite{a9}. There are also some other works related to this problem.  Hermelin etc. \cite{Hermelin45} introduced a new complexity hierarchy for parameterized problems named WK/MK-hierarchy. The lowest hardness class in the hierarchy is WK[1]. They found some problems are complete for WK[1], and conjectured that no WK[1]-complete problem admits a polynomial Turing kernelization, because if there exists a polynomial Turing kernelization for a WK[1]-complete problem, then all problems in WK[1] will have polynomial Turing kernelizations. The situations of the WK[k]-complete problems are alike for all $k \ge 2$. Jansen and Marx \cite{jansen marx} studied the $\mathcal{F}$-Subgraph Test and $\mathcal{F}$-Packing problems where $\mathcal{F}$ is a class of graphs, and pointed out which classes $\mathcal{F}$ make the two problems tractable in one of the following senses: polynomial-time solvable, admitting a polynomial many-one kernel or admitting a polynomial Turing kernel.

It is well-known that P and NP-complete are both subsets of NP. A natural question is whether there are languages between P and NP-complete.
\begin{definition}
(NP-intermediate) Under the assumption of NP $\not =$ P, languages that are in class NP but are neither in the class P nor NP-complete are called NP-intermediate.
\end{definition}

If we can find any problem in NP-intermediate without any assumption, then NP $\not =$ P. According to Ladner's Theorem \cite{ladner theorem}, if NP $\not =$ P, then there exist infinitely many NP-intermediate problems.  However, the problem constructed by Ladner is artificial, and finding natural NP-intermediate problems under the assumption NP $\not =$ P is a longstanding open problem in computational complexity community.

Graph isomorphism and integer factorization are famous natural candidate NP-intermediate problems. Sch\"{o}ning \cite{Schoning324} proved that if graph isomorphism is in NP-complete, then polynomial hierarchy will collapse to the second level. Babai \cite{babai} announced a quasipolynomial time algorithm for graph isomorphism, which is a big breakthrough on this problem and also a convincing evidence showing that graph isomorphism is not in NP-complete. Integer factorization is known to be in both NP and co-NP, so the problem in NP or coNP would imply NP $=$ coNP. However, as far as we know, it has no impact on complexity assumption if these two problems are in P.

This example is come from the Stack Exchange. Under the assumption of NEXP $\not =$ EXP. One the one hand,  NEXP $\not =$ EXP means NP $\not =$ P, and there is no sparse set in NP-complete based on Mahaney's theorem \cite{mahaney}. On the other hand, a proper padded version of NEXP-complete problem is in NP, but not in P, as that will contradict NEXP $\not =$ EXP. Moreover, the problem is a sparse set. Thus, the proper padded version of NEXP-complete problem must be in NP-intermediate. Although this result is based on a widely believed assumption, the padded versions of problems are not so natural. In addition, the NEXP-complete problems are not proverbial.

There are some natural problems which have been proved in NP-intermediate based on ETH. For example, tournament dominating set problem \cite{Megiddo}, V-C dimension \cite{downey book}, densest $k$-subgraph with perfect completeness \cite{Braverman, FeigeQP}. Although this type of candidate NP-intermediate problems are very natural, ETH is a very strong assumption.

\textbf{Our results.} In this paper, we build a new bridge between parameterized complexity and structural complexity of computational complexity, and then use this bridge to construct some new frameworks. There are two main ideas in the frameworks. The first, if a parameterized problem is still (NP-)hard when restricting the parameter to a ``small'' function (such as, logarithmic, polylogarithmic etc.) depending on the instance length, then this problem probably has no small (Turing) compression. The second, if a parameterized problem has a small compression, then we can obtain the constraint version of the parameterized problem by restricting the parameter to a ``small'' function depending on the instance length, then the constraint version problem probably cannot be a (NP-)hard one.

These frameworks have three main applications based on the assumptions that the polynomial hierarchy and the exponential hierarchy will not collapse.

(1) These frameworks can obtain the Turing compression (kernel) lower bounds of some important FPT problems by taking advantage of the fruitful research results in classic complexity theory. For example, the frameworks prove that edge clique cover has no polynomial  Turing compression unless the exponential hierarchy collapses to the second level, and has no linear Turing compression unless the polynomial hierarchy collapses to the second level. Thus, this paper solves an important open problem (which mentioned in abstract) in parameterized complexity.

(2) These frameworks can also get better compression (kernel) lower bounds for these FPT problems. For example, the frameworks also prove that edge clique cover has no $2^{o(k)}$ compression unless the polynomial hierarchy collapses to the third level (matching the result in \cite{cygan}), and has no linear compression unless NP = P. Thus, the frameworks are also new ones to deal with many-one compression lower bounds.

(3) These frameworks can find a large number of natural NP-intermediate problems by taking advantage of the fruitful research results in parameterized complexity. For example, the problem of deciding whether $2^{\sqrt{logn}}$ vertices can be deleted from a $n$-vertices graph $G$ in order to turn it into a forest is in NP-intermediate, unless exponential hierarchy collapses to the third level. In particular, if this problem is in NP-complete then polynomial hierarchy collapses to the third level, and if this problem is in P then NP $\subseteq$ DTIME$(n^{O(logn)})$. Thus, the paper makes some contributions to the longstanding open problem (which mentioned in abstract) in classic computational complexity.

\textbf{Organization.} Section 2 gives preliminaries including other definitions that are needed in this paper. Section 3 demonstrates the frameworks for solving (Turing) compression lower bound problem and finding natural candidate problems in NP-intermediate. Section 4 applies these frameworks to concrete problems. Section 5 is conclusion and final remarks.

\section{Preliminaries}
At first, we define the non-elementary tower function. Let function $p: \mathbb{N} \rightarrow \mathbb{N}$ be the non-elementary tower function, and $p(0,n)=n$, $p(k+1,n)=2^{p(k,n)}$. The inverse of $p$ would be the $log^{(k)}$ function, and $log^{(0)}n=n$, $log^{(k+1)}n=log(log^{(k)}n$). $log^*n$ equals to the least integer $k$ such that $log^{(k)}n \le 1$. We also use the O or o notations in the non-elementary tower function, for example, $p(1,o(n))=2^{o(n)}$, $p(2,O(n))=2^{2^{O(n)}}$ etc.

\subsection{Notions in Parameterized Complexity}

The running time of compression and Turing compression is a polynomial function on input size. We will extend the time complexity from a polynomial function to some function $t$, in order to get the definitions of $t$-compression and $t$-Turing compression.
\begin{definition}
\label{compresion def1}
($t$-Compression) A $t$-compression for a parameterized language $L \subseteq \Sigma^* \times \mathbb{N}$ is a $t(n)$-time Turing machine $M$ ($n$ is the size of the input), which given any instance $(x,k)$ returns an instance $x'$ such that $(x,k) \in L$ if and only if $x' \in Q$ ($Q$ is a language) and with $|x'|\leqslant 2f(k)$. We also say that parameterized language $L$ has an $f$-sized $t$-compression.
\end{definition}
We say $L$ has an $f$-sized $T$-compression if $L$ has an $f$-sized $t$-compression for some $t \in T$, $L$ has an $F$-sized $t$-compression if $L$ has an $f$-sized $t$-compression for some $f \in F$, $L$ has an $F$-sized $T$-compression if $L$ has an $f$-sized $t$-compression for some $t \in T$ and $f \in F$.

The parameter $k$ is smaller than the instance length $|x|$ in almost all important problems in parameterized complexity. Thus, without loss of generality we assume that $k\leq |x|$. The same as $t$-compression, we will give a definition of $t$-Turing compression.

\begin{definition}
($t$-Turing Compression) $L\subseteq \Sigma^* \times \mathbb{N}$ is said to have an $f$-sized $t$-Turing compression if there is a Turing machine which given an input $(x,k)$ together with an $f(k)$-oracle for a classic language $Q$ decides whether $(x,k)\in L$  in time $t(n)$, where $n=|x|+k$.
\end{definition}
We define $f$-sized $T$-Turing compression, $F$-sized $t$-Turing compression and $F$-sized $T$-Turing compression similar to those in definition \ref{compresion def1}.

For a parameterized language, the parameter can be any number. If we restrict the parameter to some special function of the instance length, then the parameterized language will turn into a new language. We named this kind of languages constraint parameterized languages.
\begin{definition}
$($Constraint Parameterized Language$)$  $L \subseteq \Sigma^* \times \mathbb{N}$ is a parameterized language. For all instance $(x,k)$ of $L$, if we restrict the parameter $k$ to $f(|x|)$, then we get a constraint parameterized language $Q$, and $Q=\{(x,f(|x|)) | (x,k) \in L\}$. We say that $Q$ is the $f$-constraint parameterized language of $L$.
\end{definition}

We say $Q$ is an $F$-constraint parameterized language of $L$ if $Q$ is an $f$-constraint parameterized language of $L$ for some $f \in F$. The instances (including yes instance and no instance) of $Q$ have the property that the parameter $k=f(|x|)$. We will not consider the instances whose formats do not satisfy this property (Turing machine is very easy to recognize the incorrect format instances). Thus, we have $\overline Q=\{(x,f(|x|)) | (x,k) \in \overline L\}$, where $\overline L$ and $\overline Q$ are the complement languages of $L$ and $Q$, respectively.

In this paper, we only consider that $f(|x|)$ is a constraint number of $k$, so we have $f(|x|) \leq |x|$ for large enough $|x|$.

\begin{definition}
 $($f-Hard Parameterized Language$)$ $L \subseteq \Sigma^* \times \mathbb{N}$ is a parameterized language. If language $Q \subseteq \Sigma^* \times \mathbb{N}$ is an $f$-constraint parameterized language of $L$ as well as $Q$ is NP-hard, then we call $L$ $f$-hard parameterized language.
\end{definition}
We define $F$-hard parameterized language similar to the definition of the $F$-constraint parameterized language. $F$ can be various kinds of function classes, such as logarithmic, polylogarithmic, fractional power etc.

\subsection{Notions in Classic Computational Complexity}
NEXP (EXP) is the set of decision problems that can be solved by a non-deterministic (deterministic) Turing machine using time $2^{n^{O(1)}}$. The exponential hierarchy (sometimes called EXPH) is a hierarchy of complexity classes, which is an exponential time analogue of the well-known polynomial hierarchy (sometimes called PH).
\begin{definition}
(Exponential Hierarchy) Exponential hierarchy is the union of the language classes $\Sigma_i^{EXP}$, where $\Sigma_i^{EXP}=NEXP^{\Sigma_{i-1}^P}$ (languages computable in nondeterministic time $2^{n^c}$ for some constant c with a $\Sigma_{i-1}^P$ oracle), and again $\Pi_i^{EXP} = coNEXP^{\Sigma_{i-1}^P}$, $\Delta_i^{P}=EXP^{\Sigma_{i-1}^P}$.
\end{definition}

It is commonly believed that PH and EXPH will not collapse.

\begin{definition}
($f$-Sparse Set) Set $S \subseteq \Sigma^* $ is $f$-sparse if $\| \{ x \in S |  |x|=n \} \| \leq f(n)$ for all but finitely many $n$. We say $S$ is $F$-sparse if $S$ is $f$-sparse for some $f \in F$.
\end{definition}

People usually call polynomial-sparse set sparse set. In some papers, people also consider that set $S \subseteq \Sigma^* $ is $f$-sparse if $\| \{ x \in S |  |x| \leq n \} \| \leq f(n)$ for all $n$. However, these two kinds of definitions have no difference when set $S$ is (quasi)polynomial-sparse or subexponential-sparse.

\begin{definition}
(C/F) An advice function is a function $f: \mathbb{N} \rightarrow \Sigma^*$. Let $C$ be a complexity class and $F$ be a class of advice functions. The class $C/F$ is the collection of all sets $A$ such that for some language $B \in C$ and some $f\in F$, $A=\{ x|(x,f(|x|))\in B \}$.
\end{definition}

\begin{definition}
(Circuit Families) A $f(n)$-size circuit family is a sequence $\{C_n\}_{n \in \mathbb{N}}$ of Boolean circuit, where $C_n$ has $n$ inputs and a single output, and its size $|C_n|\leq f(n)$ for every $n$.
\end{definition}

P/poly is a famous language class in computational complexity. And it is well-known that the following three language classes are equal. The first is P/poly, the second the class of languages which have polynomial size circuits, and the third the class of languages which are polynomial time Turing reducible to some sparse set.

\subsection{Concrete Parameterized Problems}

Next, we will give the definitions of the parameterized problems which will be employed in this paper. All these problems are in FPT.

\begin{definition}
(Edge Clique Cover) The input is an undirected graph G and a nonnegative integer k, k is the parameter, the problem asks if there exist a set of k subgraphs of G, such that each subgraph is a clique and each edge of G is contained in at least one of these subgraphs.
\end{definition}
Edge clique cover is a NP-complete problem \cite{james orlin}. It is also in PFT \cite{gramm}. In their paper \cite{cygan3456} Cygan etc. proved that it has no polynomial compression unless co-NP $\subseteq$ NP/poly and PH $= \Sigma_3^P$. Then in the paper \cite{cygan} Cygan etc. improved the lower bound to $2^{o(k)}$ under the same assumption.

Treewidth and pathwidth of a graph are two well-known parameters of structures that measure their similarity to a tree and a path, respectively. They are commonly used as parameters in the parameterized complexity. We use $tw(G)$ and $pw(G)$ to denote the treewidth and the pathwidth of graph $G$, respectively. Bounded pathwidth is a more restrictive concept compared with bounded treewith. For any graph $G$, $pw(G) \leq tw(G)$.

Given graph $G$ together with sets $L(v)\in \mathbb{N}$, one for every vertex $v$, then $G$ is $L$-colorable if there exists a coloring $c: V(G) \rightarrow \mathbb{N}$, which is proper and for every vertex $v$ we have $c(v) \in L(v)$. $G$ is called $\mathcal{C}$-choosable if for any list assignment $L: V(G) \rightarrow 2^{\mathbb{N}}$ with $|L(v)|=\mathcal{C}$ for all vertex $v$ in $G$, $G$ is $L$-colorable.
\begin{definition}
($\mathcal{C}$-Choosability) $G$ is a graph and the parameter $tw$ is the treewidth of $G$, constant $\mathcal{C} \ge3$. Deciding whether $G$ is $\mathcal{C}$-choosable.
\end{definition}
For each $\mathcal{C} \ge 3$, $\mathcal{C}$-choosability is $\Pi_2^P$-complete problem  \cite{shai}. And the problem is in FPT when parameterized by treewidth \cite{fellows453}.
\begin{definition}
($\mathcal{C}$-Choosability Deletion) $G$ is a graph and the parameter $tw$ is the treewidth of $G$, constant $\mathcal{C} \ge4$. The problem asks for the minimum number of vertices that need to be deleted to make $G$ $\mathcal{C}$-choosability.
\end{definition}
For each $\mathcal{C}\ge4$, $\mathcal{C}$-choosability deletion is $\Sigma_3^P$-complete problem \cite{marx choosability}. And the problem is in FPT when parameterized by treewidth \cite{marx choosability}.

For QBF problem, We will introduce its definition from \cite{vardi}. Quantified Boolean formula (QBF) extends propositional logic by introducing quantifiers over the Boolean domain \{0,1\}. We write formulas in QBF in prenex normal form $\psi =(Q_1p_1)(Q_2p_2)...(Q_ip_i)\psi'$, where the $Q$s are quantifiers, the $p$s are propositions, and $\psi'$ is a propositional formula, which we call the matrix. By bounding the number
of alternations, QBF can be stratified into classes $\Sigma_i^{QBF}$ and $\Pi_i^{QBF}$, where $i$ is the number of alternations, and a formula is in $\Sigma_i^{QBF}$ if its outermost quantifier is $\exists$ and contains $i$ alternations and in $\Pi_i^{QBF}$ if its outermost quantifier is $\forall$ and contains $i$ alternations. It is well-known that $\Sigma_i^{QBF}$ and $\Pi_i^{QBF}$ are the most famous $\Sigma_i^P$-complete problem and $\Pi_i^P$-complete problem, respectively. The treewidth of a QBF formula is defined as the width of the interaction graph of its CNF matrix. The interaction graph is defined with the set of propositions as vertices, and the co-occurrence (in the same clause) relation between propositions as edges.

\begin{definition}
($\Sigma_i^{QBF}$) Given a $\Sigma_i^{QBF}$ formula, constant $i \ge 1$, and the parameter $tw$ is the treewidth of the formula. Deciding whether the formula is satisfiable.
\end{definition}

Chen \cite{Hchen} proved that $\Sigma_i^{QBF}$ parameterized by treewidth is in FPT.

\begin{definition}
(Bounded Treewidth $QBF_k$) Given a QBF formula with bounded treewidth, and the parameter $k$ is the quantifier alternations number. Deciding whether the formula is satisfiable.
\end{definition}

Atserias and Oliva \cite{oliva} proved that bounded treewidth $QBF$ is PSPACE-complete. Chen \cite{Hchen} proved that bounded treewidth $QBF$ parameterized by the quantifier alternations number is in FPT.

\section{The Frameworks}
At first we will deduce some lemmas related to the frameworks. Lemma \ref{kernelization imply compression} points out that compression and Turing compression are general forms of kernelization and Turing kernelization, respectively. Lemma \ref{core theorem} and \ref{manyone core theorem} present connections between the size of (Turing) kernel of some FPT problem and the density of objective set which can be (Turing) reduced from the FPT problem.

\begin{lemma}
\label{kernelization imply compression}
L is a parameterized language.\\
(1) If L has an $f(k)$-sized $t(n)$-kernelization, then $L$ has an $f(k)$-sized $t(n)$-compression.\\
(2) If L has an $f(k)$-sized $t(n)$-Turing kernelization, then $L$ has an $f(k)$-sized $t(n)$-Turing compression.

\begin{proof}
According to the definitions, it is easy to understand that compression is a general type of kernelization. In order to simulate an $f(k)$-sized $t(n)$-kernelization of $L$, we just need to restrict the output of $f(k)$-sized $t(n)$-compression to the string of $L$.

For any instance $(x,k)$ of language $L$. Suppose $f(k)$-sized $t(n)$-Turing kernelization for $L$ is a Turing machine $M$ with $f(k)$-oracle for a parameterized language $L$. Then, whenever $M$ generates a string $(x',k')$, asks the $f(k)$-oracle for parameterized language $L$ whether $(x',k')$ in it, and the length of $(x',k')$ is bounded by $|x'|+|k'| \leq f(k)+f(k)=2f(k)$. Thus, Turing machine $M$ with $f(k)$-oracle for classic language $L$ (unparameterized version) is also an $f(k)$-sized $t(n)$-Turing compression for $L$.
\end{proof}
\end{lemma}

\begin{lemma}
\label{core theorem}
$(x,k)$ is an instance of $L \subseteq \Sigma^* \times \mathbb{N}$. $Q$ is an f-constraint parameterized language of L. Suppose $2g(f(|x|)) \leq |x|+f(|x|)$ for $|x|\ge C$, C is a larger constant. If L has a $g$-sized  $t$-Turing compression, then $Q$ is Turing reducible to a set $S \subseteq \Sigma^*$ of density $O(4^{g(f(n))})$ in time $O(nt(n))$ where $n=|x|+f(|x|)$.

\begin{proof}
Suppose $M,M_1$ are Turing machines, and $S_1,S_2 \subseteq \Sigma^*$ are classic languages. If the input $|x| \leq$C, then it is easy to prove the theorem, for the constant number is hid in the big O notation. In the following proof, we only need to deal with the long length inputs.

That $L$ has a $g$-sized $t$-Turing compression means that there exist a Turing machine $M_1$ together with a $g(k)$-oracle for $S_1$ which can decide whether $(x,k) \in L$ in time $t(|x|+k)$. Whenever $M_1$ generates a string, $M_1$ will ask the $g(k)$-oracle for $S_1$ whether the string is in it, and the length of the string is bounded by 2$g(k)$. For convenience, the Turing machine $M_1$ together with the $g(k)$-oracle for $S_1$ can be abbreviated to $M_1^{g(k)-S_1}$.

Now we consider the language $Q$. $Q$ is an $f$-constraint parameterized language of $L$, so we have $Q \subseteq L$ and $\overline Q \subseteq \overline L$, where $\overline Q$ and $\overline L$ are complement languages of $Q$ and $L$, respectively. In this way, language $Q$ can be solved by the machine $M_1^{g(f(|x|))-S_1}$. More precisely, it means that the Turing machine $M_1$ together with a $g(f(|x|))$-oracle for $S_1$ can decide whether $(x,f(k)) \in Q$ in time $t(|x|+f(|x|))$. Whenever $M_1$ generates a string, $M_1$ will ask the $g(f(|x|))$-oracle for $S_1$ whether the string is in it, and the length of the string is bounded by 2$g(f(|x|))$. Next, we will construct the set $S$ in two steps.

Step1: Running the machine $M_1^{g(f(|x|))-S_1}$ on an input $(x,f(|x|))$ of length $n$. Assume that set $S_{2,y}^{=n}$ (string $y$ is short for $(x,f(|x|))$) is the collection of the strings, which have the following properties. Firstly, $M_1$ generates the string, then asks the
$g(f(|x|))$-oracle for $S_1$ whether the string is in it, and the oracle returns yes. We define set
$S_2^{=n}=\bigcup_{|y|=n} (S_{2,y}^{=n})$=$\{ s_{2,1}^n,s_{2,2}^n,...,s_{2,t_n}^n \}$. Then we have $|s_{2,i}^n|\leq2g(f(|x|))$ $(i=1,2,...,t_n)$, so the number of elements in set $S_2^{=n}$ is equal to $t_n \leq \Sigma_{i=1}^{2g(f(|x|))}2^i=2^{2g(f(|x|))+1}$  (binary code).

Step2: Since $2g(f(|x|))\leq f(|x|)+|x|= n$ for lager $n$, creating a class of new sets $S^{=n}=\{ s_1^n,s_2^n,...,s_{t_n}^n \}$,
(where $n=C,C+1,C+2...$). $s_i^n$ is a string created by padding $n-|s_{2,i}^n|$ character ``1" on the end of the string $s_{2,i}^n$, where ``1" is a symbol not occurring in the language. It is easy to know that the number of elements in set $S^{=n}$ is equal to $\|S_2^{=n}\| \leq 2^{2g(f(|x|))+1}$. In addition, $f$ and $g$ are no decreasing functions, so $2^{2g(f(|x|))+1} \leq 2^{2g(f(|x|+f(|x|)))+1}$=$2^{2g(f(n))+1}$. Let the set $S= \bigcup_{n\geq C} S^{=n}$. Finally, we have created a set $S$ and $\|S^{=n}\|\leq 2^{2g(f(n))+1}$ .

We regard the language $Q$ as a classic language, which means that the input length is equal to the instance length plus the parameter. We will design a Turing machine $M$ with the oracle $S$ to solve $Q$. For any input string $(x,f(|x|)) \in Q$ ($n=|x|+f(|x|)$), let $M$ simulate $M_1$, except for the steps before access the oracle. Whenever $M_1$ generates a string $s$, and then asks the $g((f(|x|))$-oracle for $S_1$ whether the string is in it, $M$ will also produce the string $s$. In addition, $M$ will pad $n-|s|$ character ``1" on the end of the string $s$ to obtain a $n$-bit string $s'$, and then asks the oracle $S$ whether $s'$ in it. Based on the structure of $S$, $M_1$ asks the $g((f(|x|))$-oracle for $S_1$ for string $s$ and the oracle returns yes if and only if $M$ asks the oracle $S$ for string $s'$ and the oracle returns yes. Thus $M_1^{g(f(|x|))-S_1}$ can be simulated by $M$ with the oracle $S$. Next, we will analyse the time cost of $M$. Since the time cost of $M_1$ is $t(n)$, and the only difference between $M$ and $M_1$ are the steps before asking the oracle. Each time before $M$ asks the oracle $S$, it needs to pad at most $n$ bit ``1". Assume that $N$ refers to how many times $M$ needs to access the oracle $S$, then the time complexity for $M$ is $O(nN+t(n)) \leq O(nt(n)+t(n))=O(nt(n))$. Thus, we prove that $Q$ can also be solved by $M$ with the oracle $S$. The time complexity of $M$ is $O(nt(n))$, and $\|S^{=n}\|\leq 2^{2g(f(n))+1}$.

Combined with the input $|x| \leq$C, we can prove that if \emph{L} has a $g$-sized  $t$-Turing compression, then $Q$ is Turing reducible to a set $S\subseteq \Sigma^*$ of density $O(4^{g(f(n))})$ in time $O(nt(n))$.
\end{proof}
\end{lemma}

\begin{lemma}
\label{manyone core theorem}
$L$ is a parameterized language, and $(x,k)$ is an instance of $L$. $Q$ is an $f$-constraint parameterized language of L. Suppose $2g(f(|x|)) \leq |x|+f(|x|)$ for $|x|\ge$C, C is a larger constant. If $L$ has a $g$-sized $t$-compression, then $Q$ is many-one reducible to a set $S\subseteq \Sigma^*$ of density $O(4^{g(f(n))})$ in time $O(nt(n))$ where $n=|x|+f(|x|)$.

\begin{proof}
 Similar to lemma \ref{core theorem}. In fact, the time complexity of the reduction can be bounded by $O(n+t(n))$ with careful analysis.
\end{proof}
\end{lemma}

\subsection{Frameworks for Turing compression (kernelization) lower bound}
\begin{lemma}
\label{np in p/poly}
(see \cite{karp lipton})If every set in NP has polynomial-size family of circuits, then PH $=\Sigma_2^P$.
\end{lemma}

This lemma is the well-known Karp-Lipton theorem \cite{karp lipton}. Their original proof collapsed PH to $\Sigma_3^P$, and Michael Sipser improved it to $\Sigma_2^P$ in the same paper. After that, there has been a lot of works on the general theme inspired by the Karp-Lipton theorem \cite{arvind, bshouty, kobler}, especially, the Mahaney theorem \cite{mahaney}, which will also be used in this paper. One of the improvement on the collapse of PH was proved by Segupta, who pointed out that NP has polynomial-size family of circuits collapses the PH to $ S_2^{p} \subseteq \Sigma_2^P \cap \Pi_2^P $. Moreover, it became a stronger version of Karp-Lipton theorem after Cai \cite{cai} proved that $S_2^{p} \subseteq ZPP^{NP}$. We choose to state the results for classes in $\Sigma_i^{P}$ as these classes are well-know.

Another variant of the Karp-Lipton theorem was provided by Buhrman and Homer \cite{homer}. They proved that if NP has quasipolynomial-size circuits, then exponential hierarchy collapses to the second level. In their paper \cite{selman} Pavan et. al. improved the collapse to $S_2^{exp}$.

\begin{lemma}
\label{np in p/qpoly}
(see \cite{homer}) If any NP-complete problem is reducible to a set $S$ of density $n^{log^{k}n}$ in time $n^{log^{k}n}$, for some $k$, then EXPH $ = \Sigma_2^{EXP}$.
\end{lemma}

Actually, In their paper \cite{homer} Buhrman and Homer did not regard this result as a lemma or theorem. It is included in the proof of theorem 1 of \cite{homer}.

\begin{lemma}
\label{turing lower bound}
Suppose $L$ is an $f$-hard parameterized language and has a $g$-sized quasipolynomial-Turing compression. If there exists a constant $c > 0$, such that $g(f(n)) \le log^cn$ for all but finitely many $n$, then EXPH $ = \Sigma_2^{EXP}$.

\begin{proof}
Assume that $(x,k)$ is an instance of $L$, and NP-hard language $Q$ is an $f$-constraint parameterized language of $L$. $L$ has a g-sized quasipolynomial-Turing compression, so $L$ has a $g$-sized $t$-Turing compression for some quasipolynomial function $t$. $Q$ is Turing reducible to a set $S$ of density $O(4^{g(f(n))})$ in time $O(nt(n))$ where $n=|x|+f(|x|)$, according to lemma \ref{core theorem}. Obviously, both $O(nt(n))$ and $O(4^{g(f(n))})$ are quasipolynomial functions. According to lemma \ref{np in p/qpoly}, EXPH $ = \Sigma_2^{EXP}$.
\end{proof}
\end{lemma}

This lemma gives us a tool to get Turing compression (kernel) lower bounds under the assumption of EXPH $\not = \Sigma_2^{EXP}$. In order to obtain the compression (kernel) lower bounds of FPT languages, we only need to design polynomial-time reduction algorithms for these languages to illustrate that these languages are $f$-hard parameterized languages. Thus, this tool transforms the lower bound type problem to algorithm design type problem.

Of all the sizes of lower bounds, people pay great attention to the polynomial size. So we will give the polynomial-sized Turing kernel lower bound lemma, which is a special case of lemma \ref{turing lower bound}.

\begin{lemma}
\label{polylog imply collapse}
Suppose $L$ is a polylogarithmic-hard parameterized language. If $L$ has a polynomial-sized quasipolynomial-Turing compression, then EXPH $ = \Sigma_2^{EXP}$.

\begin{proof}
Refer to the lemma \ref{turing lower bound}. In this lemma, $f$ is a quasipolynomial function and $g$ is a polynomial function. Thus, there exists a constant $c > 0$, $g(f(n)) \le log^cn$ for all but finitely many $n$.
\end{proof}
\end{lemma}

We can also use other hypotheses to support the lower bounds of Turing kernels, such as that polynomial hierarchy will not collapse. The following lemma will give a tool for proving Turing kernel lower bounds under the assumption of PH $\not = \Sigma_2^{P}$.

\begin{lemma}
\label{linearlog imply collapse}
Suppose $L$ is an $f$-hard parameterized language and has a $g$-sized Turing compression. If there exists a constant $c > 0$, such that $g(f(n)) \le clogn$ for all but finitely many $n$, then PH $=\Sigma_2^P$.

\begin{proof}
Similar to the proof of lemma \ref{turing lower bound}. Expect for that the density of set $S$ is $O(4^{g(f(n))})=poly(n)$ and the time complexity of the Turing reduction is $poly(n)$ in this lemma. Thus, NP is polynomial time Turing reducible to a sparse set, and PH $=\Sigma_2^P$ according to the lemma \ref{np in p/poly}.
\end{proof}
\end{lemma}

\subsection{Frameworks for many-one compression (kernelization) lower bound}
A set $S$ has subexponential density if for any constant $\epsilon > 0$, $||S_{=n}|| \le 2^{n^{\epsilon}}$ for all but finitely many $n$.

\begin{lemma}
\label{mahaney theorem}
(see \cite{mahaney})  If NP is polynomial time many-one reducible to a sparse set, then NP = P.
\end{lemma}

This is Mahaney's theorem. There has been a lot of works on the variant of this theorem. One direction is to relax the reduction. One of the prominent results in this direction was proved by Ogiwara and Watanabe \cite{ogiwara}. They proved that if NP is bounded truth-table reducible to a sparse set, then NP = P. Enlarging the sparse set is another direction in which Lemma \ref{no np subexp} is one of the considerable results.

\begin{lemma}
\label{no np subexp}
(see \cite{Buhrman})  If NP is polynomial time many-one reducible to a subexponential density set, then PH $=\Sigma_3^P$.
\end{lemma}

Buhrman and Hitchcock \cite{Buhrman} proved that if NP has subexponential density under polynomial-time many-one reduction, then coNP $\subseteq$ NP/poly. Yap \cite{yap} proved that if coNP $\subseteq$ NP/poly, then polynomial hierarchy collapses to the third level, and Cai etc. \cite{cai2} improved the collapse to $\mathrm{S_2^{NP}}$. In fact, the many-one reduction in lemma \ref{no np subexp} can be generated to conjunctive reduction and query-bounded Turing reduction \cite{Buhrman}. Moreover, the following theorems concerning many-one compression (kernel) lower bounds based on the assumption of PH $\not=\Sigma_3^P$ can also be easily extended to conjunctive compression (kernel) lower bounds and query-bounded Turing compression (kernel) lower bounds \cite{open1}.

\begin{lemma}
\label{manyone lower bound s2p}
Suppose $L$ is an $f$-hard parameterized language and has a  $g$-sized compression. If for any constant $\epsilon > 0$, $g(f(n)) \le n^{\epsilon}$ for all but finitely many $n$, then PH $=\Sigma_3^P$.

\begin{proof}
Assume that $(x,k)$ is an instance of $L$, and NP hard language $Q$ is an $f$-constraint parameterized language of $L$. $L$ has a g-sized compression, so $L$ has a $g$-sized $t$-compression for some polynomial function $t$. $Q$ is many-one reducible to a set $S\subseteq \Sigma^*$ of density $O(4^{g(f(n))})$ in time $O(nt(n))$ where $n=|x|+f(|x|)$, according to lemma \ref{manyone core theorem}. Obviously, $O(nt(n))$ is a polynomial function, and $O(4^{g(f(n))}) \leq 2^{n^{\epsilon}}$ for all but finitely many $n$. Thus, $Q$ is polynomial-time many-one reducible to a set $S$ of subexponential density. According to lemma \ref{no np subexp}, PH $=\Sigma_3^P$.
\end{proof}
\end{lemma}

\begin{lemma}
\label{manyone lower bound np p}
Suppose $L$ is an $f$-hard parameterized language and has a $g$-sized compression. If there exists a constant $c > 0$, such that $g(f(n)) \le clogn$ for all but finitely many $n$, then NP = P.

\begin{proof}
Similar to lemma \ref{linearlog imply collapse}. It is easy to conclude that NP is polynomial time many-one reducible to a sparse set. Then NP = P according to lemma \ref{mahaney theorem}.
\end{proof}
\end{lemma}

\subsection{Framework for finding natural candidate NP-intermediate problems}

\begin{lemma}
\label{NPI core}
$f: \mathbb{N} \rightarrow \mathbb{N}$ is a strictly monotone increasing function. $(x,k)$ is an instance of parameterized language $L$. $Q$ is an $f$-constraint parameterized language of $L$ and $f(|x|) \in |x|^{o(1)}$.  If $L$ is an $O(|x|^{\epsilon})$-hard parameterized language, where constant $\epsilon \in (0.1]$, and $L$ has a polynomial compression, then we have\\
(1) If $Q$ is in NP-complete, then PH $=\Sigma_3^P$.\\
(2) If $Q$ is in P, then $NP \subseteq DTIME((f^{-1}(n))^{O(1)})$, where $n$ is the length of the input of the language in NP.

\begin{proof}
(1) If $Q$ is NP-complete, then $L$ is an $f$-hard parameterized language, and $L$ has a $g$-sized compression for some polynomial function $g$. We have $g(f(|x|))=poly(|x|^{o(1)})\le |x|^{\epsilon}$ for any constant $\epsilon >0$. Thus, PH $=\Sigma_3^P$ according to lemma \ref{manyone lower bound s2p}.

(2) Suppose NP-hard language $L'$ is an $O(|x|^{\epsilon})$-constraint parameterized language of $L$. Then $L'$ is an $f'$-constraint parameterized language of $L$ for some function $f'(|x|) \in O(|x|^{\epsilon})$. We will consider an instance $(x,f'(|x|)) \in L' \subseteq L$ in two steps.

Step1: Pad some special characters $\#$ on the end of string $x$ to transform the string $(x,f'(|x|)) \in L$ into $(y,f'(|x|)) \in L$, and satisfy $f'(|x|)=f(|y|)$. Thus, string $(y,f'(|x|)) \in Q \subseteq L$  (in fact, string $(y,f'(|x|))$ and string $(y,f(|y|))$ are the same), and the number of the padded characters $\#$ is less than $|y|=f^{-1}(f'(|x|))$. The time complexity of the first step is $O(|y|)$.

Step2: Consider the language $(y,f'(|x|)) \in Q$. $Q \in$ P means that we can decide it in polynomial time, so the time complexity to solve $Q$ is $(|y|+f'(|x|))^{O(1)}$.

Finally, summing up the time complexity of the two steps, we can get that $L'$ can be decided in time $(|y|+f'(|x|))^{O(1)}=(f^{-1}(f'(|x|))+f'(|x|))^{O(1)}$. It is easy to see that $f^{-1}(|x|)$ is a superpolynomial function, so we have  $(f^{-1}(f'(|x|))+f'(|x|))^{O(1)} = (f^{-1}(|x|))^{O(1)}$. $L'$ is NP-hard, so there exists a polynomial-time algorithm that, given a $n$-sized instance of any language in NP, constructs an equivalent $|x|$-sized instance of $L'$ with $|x|=n^{O(1)}$. Thus, any language with input size $n$ in NP can be solved in time $n^{O(1)}+(f^{-1}(n^{O(1)}))^{O(1)}=(f^{-1}(n))^{O(1)}$.
\end{proof}
\end{lemma}

\section{Applications}
This section falls into three parts. The first part and the second part will give Turing compression lower bounds and compression lower bounds of some FPT problems, respectively. The third part will illustrate some natural candidate problems in NP-intermediate.

At first, we need to prove some properties of the FPT problems. These problems include satisfiability for $\Sigma_i^{QBF}$ formulas parameterized by treewidth, satisfiability for bounded treewidth $QBF_k$ formulas parameterized by quantifier alternations number $k$, edge clique cover, $\mathcal{C}$-choosability parameterized by treewidth and $\mathcal{C}$-choosability deletion parameterized by treewidth.

\begin{lemma}
\label{FPT problems are hard}
The problems mentioned above have following properties. Assume that $(x,k)$ is an instance of each problem.\\
(1) Edge clique cover parameterized by clique number is an $O(log|x|)$-hard parameterized problem.\\
(2) For each constant $\mathcal{C}\geq 3$, $\mathcal{C}$-choosability parameterized by treewidth is an $O(log|x|)$-hard parameterized problem.\\
(3) For each constant $\mathcal{C}\geq 4$, $\mathcal{C}$-choosability deletion parameterized by treewidth is an $O(loglog|x|)$-hard parameterized problem.\\
(4) For each odd constant number $i \ge 3$, $\Sigma_{i}^{QBF}$ parameterized by treewidth is an $O(log^{(i-1)}|x|)$-hard parameterized problem.\\
(5) There exists a constant $\mathcal{C} \ge 1$ such that $QBF_k$ on inputs of treewidth at most $\mathcal{C}$ parameterized by quantifier alternations number $k$ is a $4log^*|x|$-hard parameterized problem.

\begin{proof}
Bounded pathwidth is a more restrictive concept than bounded treewith, so the following results are still right if replacing pathwidth with treewidth.

(1) Cygan, Pilipczuk, and Pilipczuk \cite{cygan} proved that there exists a polynomial-time algorithm that, given a 3-SAT formula with $n$ variables and $m$ clauses, constructs an equivalent edge clique cover instance $(G,k)$ with $k=O(logn)$ and $|V(G)|=O(n+m)$. Thus, edge clique cover parameterized by clique number is an $O(log|x|)$-hard parameterized problem.

(2) In the proof of the algorithmic lower bound of $\mathcal{C}$-choosability problem \cite{marx choosability}, one of the underlying results is that $\mathcal{C}$-choosability on inputs of pathwidth bounded by $O(log|x|)$ is NP-hard. More precisely,  Holyer \cite{hoyer} proved that edge 3-coloring problem is NP-complete. Marx and Mitsou \cite{marx choosability} constructed a polynomial time reduction from edge 3-coloring to (2,3)-choosability problem. Moreover, they proved that if the input instance of the reduction has $n$ vertices then the pathwidth of the output instance is bounded by $O(logn)$. Gutner and Tarsi \cite{shai} constructed a polynomial time reduction from (2,3)-choosability to $\mathcal{C}$-choosability. In addition, they proved that the pathwidth $pw(G')$ of the output instance $G'$ of the reduction is linear bounded by the pathwidth $pw(G)$ of the input instance $G$, that is, $pw(G')=O(pw(G))$.  Thus, $\mathcal{C}$-choosability problem parameterized by pathwidth is an $O(log|x|)$-hard parameterized problem.

(3) The same as the $\mathcal{C}$-choosability, in the paper of Marx and Mitsou \cite{marx choosability}, one of the underlying results is that $\mathcal{C}$-choosability deletion on inputs of pathwidth bounded by $O(loglog|x|)$ is NP-complete. More precisely, Kratochv\'{\i}l \cite{Kratochvil} proved that bipartite list 3-coloring problem is NP-hard. Marx and Mitsou \cite{marx choosability} presented a reduction from bipartite list 3-coloring to $\mathcal{C}$-choosability deletion. Moreover, they proved that if the input instance has $n$ vertices, then the pathwidth of the output instance is bounded by $O(loglogn)$. Thus, $\mathcal{C}$-choosability deletion parameterized by pathwidth is an $O(loglog|x|)$-hard parameterized problem.

(4) We can prove that $\Sigma_i^{QBF}$ problem of treewidth at most $O(log^{(i-1)}|x|)$ is still NP-hard according to Theorem 3.5 and Theorem 4.5 in \cite{vardi}. Theorem 3.5 in \cite{vardi} showed that there exists a polynomial time algorithm that, given an instance $(T,w,I)$ of tiling-check problem in NP-complete, constructs a QPTL finite model checking instance $\pi_{w,I} \models \varphi_{T,w}$. In addition, the reduction satisfies

(\rmnum{1}) $\pi_{w,I} \models \varphi_{T,w}$ if and only if $(T,w,I)$ have a solution.

(\rmnum{2}) The word model $\pi_{w,I}$ has size polynomial in $w$.

(\rmnum{3}) The formulas $\varphi_{T,w}$ are in $\Sigma_i^{QPTL}$ $(i>2)$ and are small enough so that there exists a constant $c > 0$ where $p(i-1,c|\varphi_{T,w}|)$ is in $poly(w)$.

(\rmnum{2})  and (\rmnum{3}) mean that there exists a constant $c_1 > 0$ where $|\varphi_{T,w}| \leq c_1log^{(i-1)}(|\pi_{w,I}|)$. Thus, the reduction implies that QPTL finite model checking problem is NP-hard when the formulas $\varphi_{T,w}$ are in $\Sigma_i^{QPTL}$ and $|\varphi_{T,w}| \leq c_1log^{(i-1)}(|\pi_{w,I}|)$ for some constant $c_1 > 0$.

Theorem 4.5 in \cite{vardi} proved that there exists a polynomial (precisely, $O(|\varphi||\pi|)$) time algorithm that, given a formula $\varphi$ in $\Sigma_i^{QPTL}$ (for odd $i$) and a finite word model $\pi$, constructs a formula $\psi = \varphi_{Q,\pi}$ in $\Sigma_i^{QBF}$ of size $O(|\varphi||\pi|)$ and pathwidth $|tw| \leq 2|\varphi|-1$, such that $\pi \models \varphi$ if and only if $\models \psi$.

Since $|\varphi| \leq c_1log^{(i-1)}(|\pi|)$ for some constant $c_1 > 0$, we have $|tw| \leq 2c_1log^{(i-1)}(|\pi|)-1 \leq 2c_1log^{(i-1)}(|\varphi||\pi|)-1=O(log^{(i-1)}(|\psi|))$. So $\Sigma_i^{QBF}$ is NP-hard when the pathwidth is bounded by $O(log^{(i-1)}(|\psi|))$. Finally, we prove that for each odd constant number $i \ge 3$, $\Sigma_{i}^{QBF}$ parameterized by treewidth is an $O(log^{(i-1)}|x|)$-hard parameterized problem.

(5) In \cite{oliva}, corollary 1 proved that there exists a constant $\mathcal{C} \ge 1$ such that, for every $r \ge 0$, $QBF_k$ on inputs of pathwidth at most $\mathcal{C}$ and $k=r+4log^*n$ quantifier alternations is NP-hard, where $n$ is the number of variables in the formula. It is clear that $n \leq |x|$, so bounded treewidth $QBF_k$ is NP-hard when the quantifier alternations number is $4log^*|x|$.
\end{proof}
\end{lemma}

\subsection{Turing compression (kernelization) lower bounds for concrete problems}
In this part, we will apply the frameworks of Turing compression lower bounds to solving concrete problems which have been mentioned above.

\begin{theorem}
\label{main  theorem1}
Unless EXPH = $\Sigma_2^{EXP}$,\\
(1) There is no polynomial-sized quasipolynomial-Turing compression for edge clique cover.\\
(2) For every constant $\mathcal{C} \ge 3$, there is no polynomial-sized quasipolynomial-Turing compression for $\mathcal{C}$-choosability parameterized by treewidth.\\
(3) For every constant $\mathcal{C} \ge 4$, there is no $2^{O(tw)}$-sized quasipolynomial-Turing compression for $\mathcal{C}$-choosability deletion parameterized by treewidth.\\
(4) For $\Sigma_i^{QBF}$ parameterized by treewidth, there are no $2^{O(tw)}$-sized quasipolynomial-Turing compression for it when i = 3, and no $p(i-2,o(tw))$-sized quasipolynomial-Turing compression for it when i $= 2c+3$ (constant c $\ge 1$).\\
(5) There is no $(p(\frac{k}{4}-2,1))^{O(1)}$-sized quasipolynomial-Turing compression for bounded treewidth $QBF_k$ parameterized by quantifier alternations $k$.

\begin{proof}
Lemma \ref{FPT problems are hard} proved that edge clique cover and $\mathcal{C}$-choosability (constant $\mathcal{C} \ge 3$) parameterized by treewidth are both  polylogarithmic-hard parameterized problems. Thus, the two problems both have no polynomial-sized quasipolynomial-Turing compression unless EXPH = $\Sigma_2^{EXP}$ according to lemma \ref{polylog imply collapse}.

Lemma \ref{FPT problems are hard} proved that $\mathcal{C}$-choosability deletion parameterized by treewidth is an $O(loglogn)$-hard parameterized problem for each fixed $\mathcal{C}\ge4$. There exits a function $f(n) \in O(loglogn)$ such that $\mathcal{C}$-choosability deletion is an $f$-hard parameterized problem. If there exists a $g$-sized quasipolynomial-Turing compression for $\mathcal{C}$-choosability deletion for some function $g(tw) \in 2^{O(tw)}$, then $g(f(n)) \le 2^{c_1(c_2loglogn)}=(logn)^c$ for some constant $c_1$, $c_2$ and $c=c_1+c_2$. Thus, EXPH = $\Sigma_2^{EXP}$ based on lemma \ref{turing lower bound}.

Lemma \ref{FPT problems are hard} proved that $\Sigma_i^{QBF}$ is an $O(log^{(i-1)}n)$-hard parameterized problem for each odd number $i\ge 3$. In particular, $\Sigma_3^{QBF}$ is an $O(loglogn)$-hard parameterized problem. The same as the proof of $\mathcal{C}$-choosability deletion, $\Sigma_3^{QBF}$ parameterized by treewidth has no $2^{O(tw)}$-sized quasipolynomial-Turing compression. For the case of $i\ge 5$, there exists a function $f(n) \in O(log^{(i-1)}n)$, such that $\Sigma_i^{QBF}$ is an $f$-hard parameterized problem. If there exists a $g$-sized quasipolynomial-Turing compression for $\Sigma_i^{QBF}$ for some function $g(tw) \in p(i-2,o(tw))$, then we have $g(f(n)) \le p(i-2,o(c_1log^{(i-1)}n))$ for some constant c$_1$, and it is easy to prove that $p(i-2,o(c_1log^{(i-1)}n)) \le log^cn$ for some constant c. In fact, for any  $\epsilon>0$, $p(i-2,o(c_1log^{(i-1)}n)) \le log^{\epsilon}n$. Thus, EXPH = $\Sigma_2^{EXP}$ according to lemma \ref{turing lower bound}.

Lemma \ref{FPT problems are hard} proved that bounded treewidth $QBF_k$ parameterized by quantifier alternations number $k$ is an $f$-hard parameterized problem and $f(n) = 4log^*n$. If the problem has a $g$-sized quasipolynomial-Turing compression for some function $g(k) \in (p(\frac{k}{4}-2,1))^{O(1)}$, then $g(f(n)) \leq (p(\frac{4log^*n}{4}-2,1))^c = (log(p(log^*n-1,1)))^c \leq (logn)^c$ for some constant c. Thus, EXPH = $\Sigma_2^{EXP}$ based on lemma \ref{turing lower bound}.
\end{proof}
\end{theorem}

\begin{theorem}
\label{PH to second}
Unless PH = $\Sigma_2^P$,\\
(1) There is no linear-sized Turing compression for edge clique cover.\\
(2) For every constant $\mathcal{C} \ge 3$,  there is no linear-sized Turing compression for $\mathcal{C}$-choosability parameterized by treewidth.\\
(3) For every constant $\mathcal{C} \ge 4$, there is no $2^{o(tw)}$-sized Turing compression for $\mathcal{C}$-choosability deletion parameterized by treewidth.\\
(4) For each odd constant number $i \ge 3$, there is no $p(i-2,o(tw))$-sized Turing compression for $\Sigma_i^{QBF}$ parameterized by treewidth.\\
(5) There is no $O(p(\frac{k}{4}-2,1))$-sized Turing compression for bounded treewidth $QBF_k$ parameterized by quantifier alternations $k$.

\begin{proof}
Refer to the theorem \ref{main theorem1}. The proofs of the two theorems are very similar, except for that this proof will be based on lemma \ref{linearlog imply collapse} instead of lemma \ref{turing lower bound} in the proof of theorem \ref{main theorem1}.
\end{proof}
\end{theorem}

\subsection{Better many-one compression (kernelization) lower bounds for concrete problems}
This part will give compression lower bounds for these problems which have been mentioned above. In their paper \cite{cygan}, Cygan etc. have also proved proposition (1) in theorem \ref{main compression results}, and we can obtain the same result by employing our framework. In their paper \cite{gramm}  Gramm etc. pointed out that there is a $2^{O(k)}$-sized kernel for this problem, so the kernel lower bound of edge clique cover is essentially tight.

\begin{theorem}
\label{main compression results}
Unless PH = $\Sigma_3^P$,\\
(1) (also see \cite{cygan}) There is no $2^{o(k)}$-sized compression for edge clique cover.\\
(2) For every constant $\mathcal{C} \ge 3$, there is no $2^{o(tw)}$-sized compression for $\mathcal{C}$-choosability parameterized by treewidth.\\
(3) For every constant $\mathcal{C} \ge 4$, there is no $2^{2^{o(tw)}}$-sized compression for $\mathcal{C}$-choosability deletion parameterized by treewidth.\\
(4) For each odd constant number $i \ge 3$, there is no $p(i-1,o(tw))$-sized compression for $\Sigma_i^{QBF}$ parameterized by treewidth.\\
(5) There is no $(p(\frac{k}{4}-1,1))^{o(1)}$-sized compression for bounded treewidth $QBF_k$ parameterized by quantifier alternations $k$.

\begin{proof}
Lemma \ref{FPT problems are hard} proved that edge clique cover and $\mathcal{C}$-choosability parameterized by treewidth are $f$-hard parameterized languages for some function $f(n) \in O(logn)$. If these problems have $g$-sized compression for some function $g(k) \in 2^{o(k)}$, then $g(f(n))=2^{o(logn)} \leq 2^{\epsilon logn}=n^{\epsilon}$ for any constant $\epsilon > 0$. Thus PH = $\Sigma_3^P$ according to lemma \ref{manyone lower bound s2p}. The proof of $\mathcal{C}$-choosability deletion is similar.

Lemma \ref{FPT problems are hard} proved that, for each odd number $i \ge 3$, $\Sigma_i^{QBF}$ parameterized by treewidth is an $f$-hard parameterized problem for some function $f(n) \in O(log^{(i-1)}n)$. If the problem has a $g$-sized compression for some function $g(tw) \in p(i-1,o(tw))$, then we have $g(f(n))=p(i-1,o(f(n)))=p(i-1,o(log^{(i-1)}n))\leq n^{\epsilon}$ for any constant $\epsilon > 0$. Thus PH $=\Sigma_3^P$ according to lemma \ref{manyone lower bound s2p}.

Lemma \ref{FPT problems are hard} proved that bounded treewidth $QBF_k$ parameterized by quantifier alternations number is an $f$-hard parameterized problem and $f(n)=4log^*n$. If the problem has a $g$-sized compression for some function $g(k) \in (p(\frac{k}{4}-1,1))^{o(1)}$, then $g(f(n))= (p(\frac{4log^*n}{4}-1,1))^{o(1)} \leq n^{o(1)} \leq n^{\epsilon}$ for any constant $\epsilon$. Thus, PH $=\Sigma_3^P$ according to lemma \ref{manyone lower bound s2p}.
\end{proof}
\end{theorem}

\begin{theorem}
\label{main compression results NP P}
Unless NP = P,\\
(1) There is no linear-sized compression for edge clique cover.\\
(2) For every constant $\mathcal{C} \ge 3$,  there is no linear-sized compression for $\mathcal{C}$-choosability parameterized by treewidth.\\
(3) For every constant $\mathcal{C} \ge 4$, there is no $2^{o(tw)}$-sized compression for $\mathcal{C}$-choosability deletion parameterized by treewidth.\\
(4) For each odd constant number $i \ge 3$, there is no $p(i-2,o(tw))$-sized compression for $\Sigma_i^{QBF}$ parameterized by treewidth.\\
(5) There is no $O(p(\frac{k}{4}-2,1))$-sized compression for bounded treewidth $QBF_k$ parameterized by quantifier alternations $k$.

\begin{proof}
The proof is similar to the proof of theorem \ref{PH to second}, except for that this proof will be based on lemma \ref{manyone lower bound np p} instead of lemma \ref{linearlog imply collapse} in the proof of theorem \ref{PH to second}.
\end{proof}
\end{theorem}

Another meaning of theorem  \ref{main compression results NP P} is that these problems (except for the bounded treewidth QBF) can be compressed in polynomial time if and only if they can be solved in polynomial time, because NP = P implies PH = P, and all the problems discussed in this theorem are in PH.

\subsection{Natural candidate problems in NP-intermediate}
In this part, we will use the lemma \ref{NPI core} to find out a large number of natural problems in NP-intermediate under the assumption of EXPH $\not = \Sigma_3^{EXP}$. In fact, almost all of the NP-complete parameterized problems which have polynomial kernel will conform to the lemma \ref{NPI core}.

At first, let us give the kernels size of these problems. Feedback vertex set (FVS) has a $4k^2$ kernel \cite{Thomasse}. Minimum fill-in (also known as  chordal graph completion and minimum triangulation) has a $2k^2+2k$ kernel \cite{Natanzon34}. Cluster editing has a $2k$ kernel \cite{Chen and Meng}. Clique partition has a $k^2$ kernel \cite{Mujuni and Rosamond}. Max leaf in directed graph and undirected graph has an $O(k^3)$ kernel and an $O(k)$ kernel  \cite{Daligault235}, respectively.
\begin{theorem}
\label{NPI concrete problems1}
Given a $n$-vertex graph G and a constant number $\epsilon \in (0,1)$. Unless EXPH $ = \Sigma_3^{EXP}$, the following problems are in NP-intermediate,\\
(1) (FVS) Whether $2^{log^{\epsilon}n}$ vertices can be deleted from G in order to turn it into a forest.\\
(2) (Minimum Fill-In) Whether $2^{log^{\epsilon}n}$ edges can be added to G in order to turn it into a chordal graph.\\
(3) (Cluster Editing) Whether we can transform $G$, by deleting or adding $2^{log^{\epsilon}n}$ edges, into a graph that consists of a disjoint union of cliques.\\
(4) (Clique Partition) Whether the edge-set of G can be partitioned into $2^{log^{\epsilon}n}$ cliques.\\
(5) ((Directed) Max Leaf) Whether (directed graph) G has a spanning tree with $2^{log^{\epsilon}n}$ leaves.\\
More precisely, if these problems are in NP-complete then PH $= \Sigma_3^P$, and if these problems are in P then NP $\subseteq DTIME(m^{polylogm})$, where $m$ is the length of the input of the language in NP.

\begin{proof}
Firstly, let us consider the first problem, and we name it $2^{log^{\epsilon}n}$-constraint FVS.

Suppose function $f(z)=2^{log^{\epsilon}z}$, the derivation of $f$ is $f'(z)=\frac{\epsilon f(z)}{zlog^{1-\epsilon}z}$, where $\epsilon \in (0,1)$. Thus, $f'(z)>0$ for all $z>1$ and $f$ is a strictly monotone increasing function. We obtain the inverse function of $f$ by adding $log$ functions to both sides of the equation, then we have $z=2^{log^{\frac{1}{\epsilon}}(f(z))}$, and $f^{-1}(z)=2^{log^{\frac{1}{\epsilon}}z}$. In addition, constant $\epsilon \in (0,1)$ implies $(f^{-1}(z))^{O(1)}=2^{O(log^{\frac{1}{\epsilon}}z)}=z^{polylogz}$.

Suppose that $(x,k)$ is an instance of FVS.  FVS has an $O(k^2)$ kernel, so there exists a polynomial-time reduction that, given an FVS instance $(x,k)$, constructs a new FVS instance $(x',k')$ with $|x'|=O(k'^c)$ for some constant c. Thus FVS is still NP-hard when the parameter $k'=O(|x'|^{\frac{1}{c}})$. FVS is an $O(n^{\epsilon_1})$-hard parameterized language, where $\epsilon_1$ is some constant and $\epsilon_1 \in (0,1]$.

Since $2^{log^{\epsilon}n} \leq 2^{log^{\epsilon}|x|}= |x|^{\frac{1}{log^{1-\epsilon}|x|}}$ ,it is clear that $2^{log^{\epsilon}n}$-constraint FVS is a $|x|^{o(1)}$-constraint parameterized language of FVS.

According to lemma \ref{NPI core}, if $2^{log^{\epsilon}n}$-constraint FVS is in NP-complete then polynomial hierarchy will collapse to $\Sigma_3^P$, and if the problem is in P then NP $\subseteq$ DTIME$(m^{polylogm})$. On the one hand, PH $=\Sigma_3^P$ implies EXPH $ = \Sigma_3^{EXP}$ (padding argument). On the other hand, the Proposition 2 in \cite{homer} proved that NP $\subseteq$ DTIME$(m^{polylogm})$ implies NEXP = EXP. If NEXP = EXP then EXPH  = NEXP \cite{Dawar,Mocas}, it also means EXPH $ = \Sigma_3^{EXP}$. Thus, if $2^{log^{\epsilon}n}$-constraint FVS is in NP-intermediate then EXPH $ = \Sigma_3^{EXP}$.

The proofs of other problems are similar to the proof of the first problem.
\end{proof}
\end{theorem}
Besides $f(z)=2^{log^{\epsilon}z}$, we can also choose other functions which conform to lemma \ref{NPI core}.

In the function $f(z)=2^{log^{\epsilon}z}$, $\epsilon$ is a constant. We can choose any constant $0<\epsilon <1$ for the function. For example, if we choose $\epsilon=0.5$, then we have the following corollary.
\begin{corollary}
\label{NPI concrete problems2}
Given a $n$-vertex graph G. Unless EXPH $ = \Sigma_3^{EXP}$, the following problems are in NP-intermediate,\\
(1) (FVS) Whether $2^{\sqrt{logn}}$ vertices can be deleted from G in order to turn it into a forest.\\
(2) (Minimum Fill-In) Whether $2^{\sqrt{logn}}$ edges can be added to G in order to turn it into a chordal graph.\\
(3) (Cluster Editing) Whether we can transform $G$, by deleting or adding $2^{\sqrt{logn}}$  edges, into a graph that consists of a disjoint union of cliques.\\
(4) (Clique Partition) Whether the edge-set of G can be partitioned into $2^{\sqrt{logn}}$  cliques.\\
(5) ((Directed) Max Leaf) Whether (directed graph) G has a spanning tree with $2^{\sqrt{logn}}$ leaves.\\
More precisely, if these problems are in NP-complete then PH $= \Sigma_3^P$, and if these problems are in P then NP $\subseteq DTIME(m^{O(logm)})$, where $m$ is the length of the input of the language in NP.
\end{corollary}

Besides the problems mentioned above, there are many other problems that have similar results, such as, Vertex Cover \cite{downey book}, Disjoint Triangles \cite{Fellows triangle}, 3-Hitting Set \cite{Abu-Khzam}, (Directed) Max Internal Spanning Tree \cite{Gutin, Wenjun Li}  etc.
\section{Conclusion and final remarks}
This paper found a new connection between parameterized complexity and structural complexity of classic computational complexity. Then, we created some frameworks with this connection. Moreover, These frameworks can not only obtain Turing compression (kernelization) lower bounds of some important FPT problems, but also find a large number of natural candidate NP-intermediate problems.

Similar to the definition of Turing compression (kernelization), we define conondeterministic Turing compression (kernelization) by changing the deterministic Turing machine into the conondeterministic Turing machine. It is not hard to obtain the conondeterministic Turing compression (kernelization) lower bound according to the same methodology in this paper as well as the assumption of coNP $\not \subseteq$ NP/qpoly \cite{selman}.

We believe that the frameworks could get (Turing) kernel lower bounds of other important problems. Besides, maybe we could obtain other interesting results in parameterized complexity and classic computational complexity by employing the frameworks.

\textbf{Acknowledgement:} First and foremost, the author thanks Jianer Chen for helpful discussions and advice, and Marek Cygan for the helpful e-mail. Besides, the author thanks Zhipeng Tang, Changhui Wang, Jianxin Wang, Guangwei Wu, Chao Xu for a lot of help during research and study. Last but not least, the author thanks Qilong Feng and Feng Shi for the useful reference material.

\begin{figure}[htbp]
\centering
\includegraphics[scale=0.35]{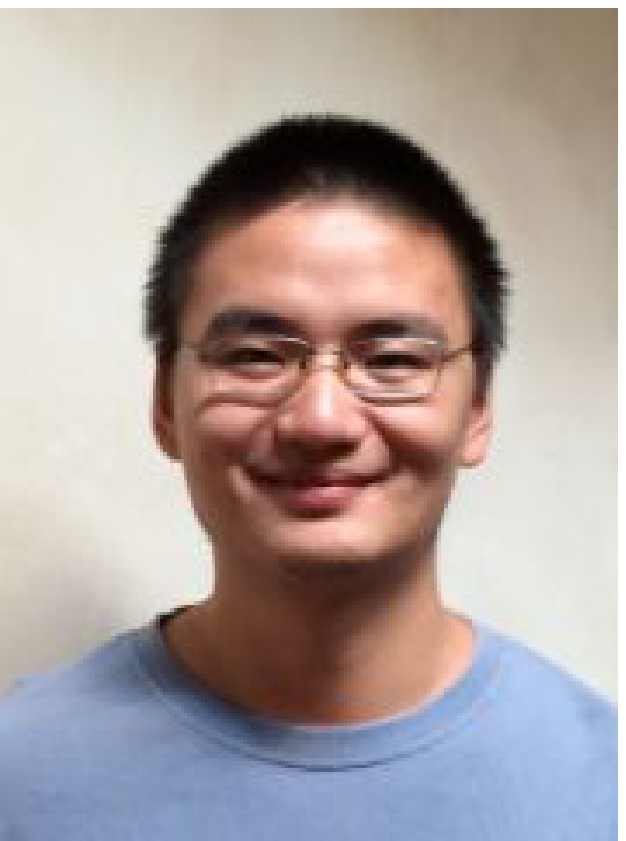}
\caption{Weidong Luo received the B.S. degree in information security from Central South University, China, in 2013. Then he  obtained his master degree in computer science from Central South University, China, in 2016. His research interests include kernelization, parameterized algorithm, structural complexity.}\label{Fig}
\end{figure}

\end{document}